\documentclass[ ]{aa}
\usepackage{graphicx}
\usepackage{times}
\newcommand{\srm}{\scriptscriptstyle \rm}
\newcommand{\bgal}{b_{\srm II}}

\newcommand{\bft}{\rm}
\begin{document}
\title{Confirmation of the existence of coherent orientations of quasar 
  polarization vectors on cosmological scales\thanks{Based in part on 
  observations collected at the European Southern Observatory (ESO, La Silla)}}
\titlerunning{Coherent orientations of quasar polarization vectors}
\author{D. Hutsem\'ekers\inst{1,}\inst{2,}\thanks{Also, Chercheur 
  Qualifi\'e au Fonds National de la Recherche Scientifique (FNRS, Belgium)} 
  \and  H. Lamy\inst{2}}
\institute{European Southern Observatory, Casilla 19001, Santiago 19,
Chile \and Institut d'Astrophysique, Universit\'e de Li\`ege,  5 av. de 
Cointe, B-4000 Li\`ege, Belgium}
\date{Received ; accepted }

\abstract{
In order to verify the existence of coherent orientations of quasars
polarization vectors on very large scales, we have obtained new
polarization measurements for a sample of quasars located in a given
region of the three-dimensional Universe where the range of
polarization position angles was predicted in advance.
{\bft For this new sample, the hypothesis of uniform distribution of
polarization position angles may be rejected at the 1.8\% significance
level on the basis of a simple binomial test.  This result provides an
independent confirmation of the existence of alignments of quasar
polarization vectors on very large scales.  In total, out of 29
polarized quasars located in this region of the sky, 25 have their
polarization vectors coherently oriented. This alignment occurs at
redshifts $z \simeq$ 1-2 suggesting the presence of correlations in
objects or fields on Gpc scales. More global statistical tests applied
to the whole sample of polarized quasars distributed all over the sky
confirm that polarization vectors are coherently oriented in a few
groups of 20-30 quasars.
Some constraints on the phenomenon are also derived. Considering more
particularly the quasars in the selected region of the sky, we found
that their polarization vectors are roughly parallel to the plane of
the Local Supercluster. But the polarization vectors of objects along
the same line of sight at lower redshifts are not accordingly aligned.
We also found that the known correlations between quasar intrinsic
properties and polarization are not destroyed by the alignment
effect.}
Several possible mechanisms are discussed, but the interpretation of
this orientation effect remains puzzling.
}

\maketitle

\section{Introduction}

Considering a sample of 170 optically polarized quasars with accurate
polarization measurements, we recently found that quasar polarization
vectors are not randomly oriented on the sky as naturally
expected. Indeed, in some regions of the three-dimensional Universe
{\bft (i.e. in regions delimited in right ascension, declination, and
redshift)}, the quasar polarization position angles appear
concentrated around preferential directions, suggesting the existence
of large-scale coherent orientations --or alignments-- of quasar
polarization vectors (Hutsem\'ekers 1998, hereafter Paper~I).

Mainly because the polarization vectors of objects located along the
same line of sight but at different redshifts are not accordingly
aligned, possible instrumental bias and contamination by interstellar
polarization are unlikely to be responsible for the observed effect
(cf. Paper~I for a more detailed discussion).  The very large scale at
which these coherent orientations are seen suggests the presence of
correlations in objects or fields on spatial scales $\sim 1000 \,
h^{-1}$ Mpc at redshifts $z \simeq$ 1--2, possibly unveiling a new
effect of cosmological importance.

Although we found this orientation effect statistically significant,
the sample is not very large, and further investigation is needed to
confirm it, especially in view of its very unexpected nature. One of
the simplest tests consists in measuring the polarization of a new
sample of quasars located in one of the regions of the sky where an
alignment was previously found and where we can predict in advance the
range of polarization position angles.  Such polarimetric observations
have been recently carried out, and the results are presented here.

In Sect.~2, we present the new polarimetric observations, as well as a
compilation of the most recent measurements from the literature.  The
results of the statistical analysis --confirming the orientation
effect-- are given in Sect.~3. Then, from the properties of the
objects in the region of alignment we derive some constraints on
possible interpretations.  These are discussed in Sect.~4. Conclusions
form the last section.

\section{New observations and compilation of data}

In Paper~I, we identified a region in the sky (region A1) where
nearly all quasar polarization position angles lie in the range
146$\degr$ -- 46$\degr$.  This region is delimited in right ascension
by $11^{h}15^{m} \leq \alpha \leq 14^{h}29^{m}$ and in redshift by
$1.0 \leq z \leq 2.3$.

A new sample of quasars\footnote{All throughout this paper we use
``quasar'' without any distinction between quasars and quasi-stellar
objects (QSOs)} located in this region was therefore selected, mainly
from the quasar catalogues of V\'eron-Cetty \& V\'eron (1998) and
Hewitt \& Burbidge (1993). This sample was observed during two runs at
ESO La Silla in 1998 and 1999, using the ESO 3.6m telescope equipped
with EFOSC2 in its polarimetric mode.  In order to minimize the
contamination by interstellar polarization in our Galaxy, only objects
at high galactic latitudes $|\bgal| \geq 35\degr$ were considered.
The selection of the targets at the telescope was not random: the
brightest objects were given higher priority as well as objects at the
center of the alignment region where the orientation effect is
suspected to be stronger.  Also, radio-loud and broad absorption line
(BAL) quasars were preferred since these objects are more likely to be
significantly polarized (Impey \& Tapia 1990, Hutsem\'ekers et
al. 1998, Schmidt and Hines 1999). In this view, 3 BAL quasars
recently discovered by Brotherthon et al. (1998) were added to the
sample. At the end, polarimetric data were secured for 28 quasars
belonging to region A1, with a typical uncertainty of 0.2\% on the
polarization degree.  About half of them appear significantly
polarized. These data are presented in Lamy \& Hutsem\'ekers (2000)
with full account of the observation and reduction details.

\begin{table}
\caption[ ]{The additional sample of polarized quasars}
\begin{tabular}{lllrlrrr}
\hline \\[-0.10in]
Object  & $\bgal$   & $z$ & \  $p$ \ \ \ & \ $\sigma_{p}$ & $\theta$ \ & $\sigma_{\theta}$ & Ref \\
(B1950) & ($\degr$) &     & \ \ \ (\%) \ &   (\%)         & ($\degr$)  & ($\degr$)         &     \\[0.05in] 
\hline \\[-0.10in]
 B0004$+$017 & $-$59 & 1.711 &   1.29 & 0.28 &  122 &   6 &  8 \\
 B0010$-$002 & $-$61 & 2.145 &   1.70 & 0.77 &  116 &  13 &  8 \\
 B0025$-$018 & $-$64 & 2.076 &   1.16 & 0.52 &  109 &  13 &  8 \\
 B0046$-$315 & $-$86 & 2.721 &  13.30 & 2.00 &  159 &   4 &  7 \\
 B0109$-$014 & $-$64 & 1.758 &   1.77 & 0.35 &   76 &   6 &  8 \\
 B0117$-$180 & $-$79 & 1.790 &   1.40 & 0.46 &   13 &  10 &  8 \\
 B0226$-$104 & $-$62 & 2.256 &   2.51 & 0.25 &  165 &   3 &  8 \\
 B0422$-$380 & $-$45 & 0.782 &   6.20 & 3.00 &  173 &  14 &  7 \\
 B0448$-$392 & $-$40 & 1.288 &   2.90 & 1.00 &   49 &  10 &  7 \\
 B0759$+$651 & $+$32 & 0.148 &   1.45 & 0.14 &  119 &   3 &  8 \\
 B0846$+$156 & $+$33 & 2.910 &   0.80 & 0.21 &  151 &   8 &  9 \\
 B0856$+$172 & $+$36 & 2.320 &   0.70 & 0.24 &    0 &  10 &  9 \\
 B0932$+$501 & $+$47 & 1.914 &   1.39 & 0.16 &  166 &   3 &  8 \\
 B1009$+$023 & $+$44 & 1.350 &   0.77 & 0.19 &  137 &   7 &  9 \\
 B1051$-$007 & $+$50 & 1.550 &   1.90 & 0.19 &   90 &   3 &  9 \\
 B1157$-$239 & $+$37 & 2.100 &   1.33 & 0.17 &   95 &   4 &  9 \\
 B1157$+$014 & $+$61 & 1.990 &   0.76 & 0.18 &   39 &   7 &  9 \\
 B1203$+$155 & $+$74 & 1.630 &   1.54 & 0.20 &   30 &   4 &  9 \\
 B1205$+$146 & $+$74 & 1.640 &   0.83 & 0.18 &  161 &   6 &  9 \\
 B1215$+$127 & $+$73 & 2.080 &   0.62 & 0.24 &   17 &  12 &  9 \\
 B1216$-$010 & $+$61 & 0.415 &   6.90 & 0.80 &    8 &   3 &  7 \\
 B1219$+$127 & $+$74 & 1.310 &   0.68 & 0.20 &  151 &   9 &  9 \\
 B1222$-$016 & $+$60 & 2.040 &   0.80 & 0.22 &  119 &   8 &  9 \\
 B1235$-$182 & $+$44 & 2.190 &   1.02 & 0.18 &  171 &   5 &  9 \\
 B1239$+$099 & $+$72 & 2.010 &   0.82 & 0.18 &  161 &   6 &  9 \\
 B1256$-$220 & $+$41 & 1.306 &   5.20 & 0.80 &  160 &   4 &  7 \\
 B1256$-$175 & $+$45 & 2.060 &   0.91 & 0.19 &   71 &   6 &  9 \\
 B1302$-$102 & $+$52 & 0.286 &   1.00 & 0.40 &   70 &  11 &  7 \\
 B1305$+$001 & $+$62 & 2.110 &   0.70 & 0.22 &  151 &   9 &  9 \\
 B1333$+$286 & $+$80 & 1.910 &   5.88 & 0.20 &  161 &   1 &  9 \\
 B1402$+$436 & $+$68 & 0.324 &   7.55 & 0.22 &   33 &   1 &  8 \\
 B1443$+$016 & $+$52 & 2.450 &   1.33 & 0.23 &  159 &   5 &  9 \\
 B1452$-$217 & $+$33 & 0.773 &  12.40 & 1.50 &   60 &   3 &  7 \\
 B1500$+$084 & $+$54 & 3.940 &   1.15 & 0.33 &  100 &   9 &  9 \\
 B1524$+$517 & $+$52 & 2.873 &   2.71 & 0.34 &   94 &   4 &  8 \\
 B1556$+$335 & $+$50 & 1.650 &   1.31 & 0.47 &   70 &  10 &  8 \\
 B2115$-$305 & $-$44 & 0.980 &   3.40 & 0.40 &   67 &   3 &  7 \\
 B2118$-$430 & $-$45 & 2.200 &   0.66 & 0.20 &  133 &   9 &  9 \\
 B2128$-$123 & $-$41 & 0.501 &   1.90 & 0.40 &   64 &   6 &  7 \\
 B2135$-$147 & $-$43 & 0.200 &   1.10 & 0.40 &  100 &  10 &  7 \\
 B2201$-$185 & $-$51 & 1.814 &   1.43 & 0.51 &    7 &  10 &  8 \\
 B2341$-$235 & $-$74 & 2.820 &   0.64 & 0.20 &  122 &   9 &  9 \\
 B2358$+$022 & $-$58 & 1.872 &   2.12 & 0.51 &   45 &   7 &  8 \\
\hline \\[-0.10in]
 B0019$+$011 & $-$61 & 2.124 &   0.76 & 0.19 &   26 &   7 &  8 \\
 B0059$-$275 & $-$88 & 1.590 &   1.45 & 0.23 &  171 &   5 &  9 \\
 B0146$+$017 & $-$58 & 2.909 &   1.23 & 0.21 &  141 &   5 &  8 \\
 B0946$+$301 & $+$50 & 1.216 &   0.85 & 0.14 &  116 &   5 &  8 \\
 B1011$+$091 & $+$49 & 2.266 &   1.54 & 0.23 &  136 &   4 &  8 \\
 B1151$+$117 & $+$69 & 0.180 &   0.72 & 0.18 &  100 &   7 &  9 \\
 B1246$-$057 & $+$57 & 2.236 &   1.96 & 0.18 &  149 &   3 &  8 \\
 B1413$+$117 & $+$65 & 2.551 &   2.53 & 0.29 &   53 &   3 &  8 \\
 B2240$-$370 & $-$61 & 1.830 &   2.10 & 0.19 &   28 &   3 &  9 \\
\hline\end{tabular}
\footnotesize{References: (7) Visvanathan \& Wills 1998, (8) 
Schmidt \& Hines 1999, (9) Lamy \& Hutsem\'ekers 2000}
\vspace*{0.5cm}
\end{table}

In the meantime, two major quasar polarimetric surveys --obviously not
restricted to region A1-- have been published by Visvanathan \& Wills
(1998) and by Schmidt and Hines (1999). Several of their targets are
located in region A1, but most of them are unfortunately redundant with
ours and generally measured with less accuracy. All these new data have
been compiled, also including a few additional measurements we did
ourselves in the framework of a polarimetric study of radio-loud BAL
quasars (Hutsem\'ekers \& Lamy 2000).

The new polarimetric data are summarized in Table~1. They refer to
quasars distributed all over the sky, which complement the sample of
170 polarized quasars studied in Paper~I. The 43 objects reported in
the first part of Table~1 are significantly polarized and fulfil the
criteria defined in Paper~I: $p \geq 0.6\% $, $\sigma_{\theta} \leq
14\degr$, and $|\bgal| \geq 30\degr$, where $p$ is the polarization
degree and $\sigma_{\theta}$ the uncertainty of the polarization
position angle $\theta$. These constraints ensure that most objects
are significantly and intrinsically polarized with little
contamination by the Galaxy, and that the polarization position angles
are measured with a reasonable accuracy (Paper~I). If an object has
been observed more than once, only the best value is kept i.e. the
measurement with the smallest uncertainty $\sigma_{p}$ on the
polarization degree. Let us recall that our previous sample of 170
objects was similarly selected from 525 measurements compiled from the
literature. The second part of Table~1 lists quasars already studied
in Paper~I, and for which better data (i.e. with a smaller
$\sigma_{p}$) have been obtained in the recent surveys. Note that
these new measurements are in good agreement with the older ones, as
well as measurements obtained by different authors, providing
confidence in the quality of the data.

Together with the data from Paper~I, the total sample of polarized
quasars then amounts to 213 objects distributed all over the sky.

\section{Statistical analysis and results}

\begin{figure}[t]
\resizebox{\hsize}{!}{\includegraphics*{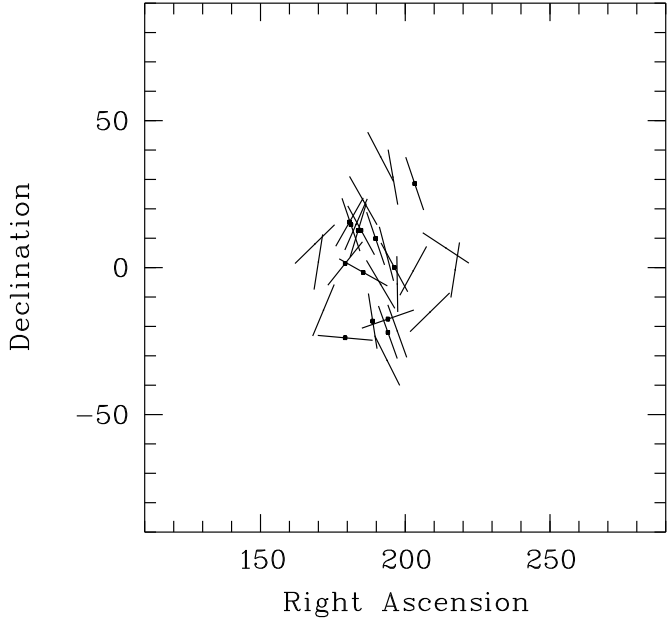}}
\caption[ ]{A map of the polarization vectors of all significantly
polarized ($p \geq 0.6 \%$ and $\sigma_{\theta} \leq 14\degr$) quasars
with right ascensions $11^{h}15^{m} \leq \alpha \leq 14^{h}29^{m}$,
and redshifts $1.0 \leq z \leq 2.3$. The vector length is arbitrary.
The 13 new objects are indicated by additional points} 
\label{map}
\end{figure}

We first want to test the hypothesis that the polarization position
angles of quasars located in region A1 preferentially lie in the
interval [146$\degr$ -- 46$\degr$] instead of being uniformly
distributed. This angular sector was selected prior to the new
observations --on the basis of the results of Paper~I--, and the
polarization position angles have been measured for a sample of
quasars different from that one at the origin of the detection of the
effect.  Out of the 13 new significantly polarized quasars in region
A1 (Table~1), 10 have their polarization position angles in the
expected range. To test the null hypothesis $H_0$ of uniform
distribution of circular observations against the alternative of
sectoral preference, we may use a simple binomial test (e.g. Lehmacher
\& Lienert 1980, Siegel 1956).  {\bft If $P_{\srm A}$ is the
probability under $H_0$ that a polarization position angle falls in
the angular sector [146$\degr$ -- 46$\degr$], then $P_{\srm A} =
80\degr/180\degr$. If $L$ denotes the number of polarization position
angles falling in [146$\degr$ -- 46$\degr$], $L$ has a binomial
distribution under $H_0$, such that the probability to have
$L_{\star}$ or more polarization angles in [146$\degr$ -- 46$\degr$] is
\begin{equation} 
P(L \geq L_{\star}) = \sum_{l=L_{\star}}^{N} 
\left(\begin{array}{c}N\\l\end{array}\right)
P_{\srm A}^{l} \; (1-P_{\srm A})^{N-l}\, . 
\end{equation}
With $N=13$ and $L_{\star}=10$, we compute $P(L \geq 10)$ =
1.8\%. This indicates that the hypothesis of uniform distribution of
polarization position angles may be rejected at the 1.8\% significance
level in favour of coherent orientation.}

A map of the quasar polarization vectors is illustrated in Fig.~1,
including the objects from Paper~I. An alignment is clearly seen, with
a net clustering of polarization vectors around $\theta \sim 165\degr
- 170\degr$. Altogether, there are 29 significantly polarized quasars
in this region, and 25 of them have their polarization vectors aligned
i.e. their polarization position angles in the range 146$\degr$ --
46$\degr$ (Tables~1 and 2, and Paper~I).  It is interesting to note
that the effect is stable when we increase the polarization degree
cutoff (then decreasing the probability of a possible contamination):
out of 22 quasars with $p \geq 0.8\%$, 19 have their polarization
vectors aligned, and out of 17 quasars with $p \geq 1.0\%$, 16 have
their polarization vectors aligned.

\begin{figure}[t]
\resizebox{\hsize}{!}{\includegraphics*{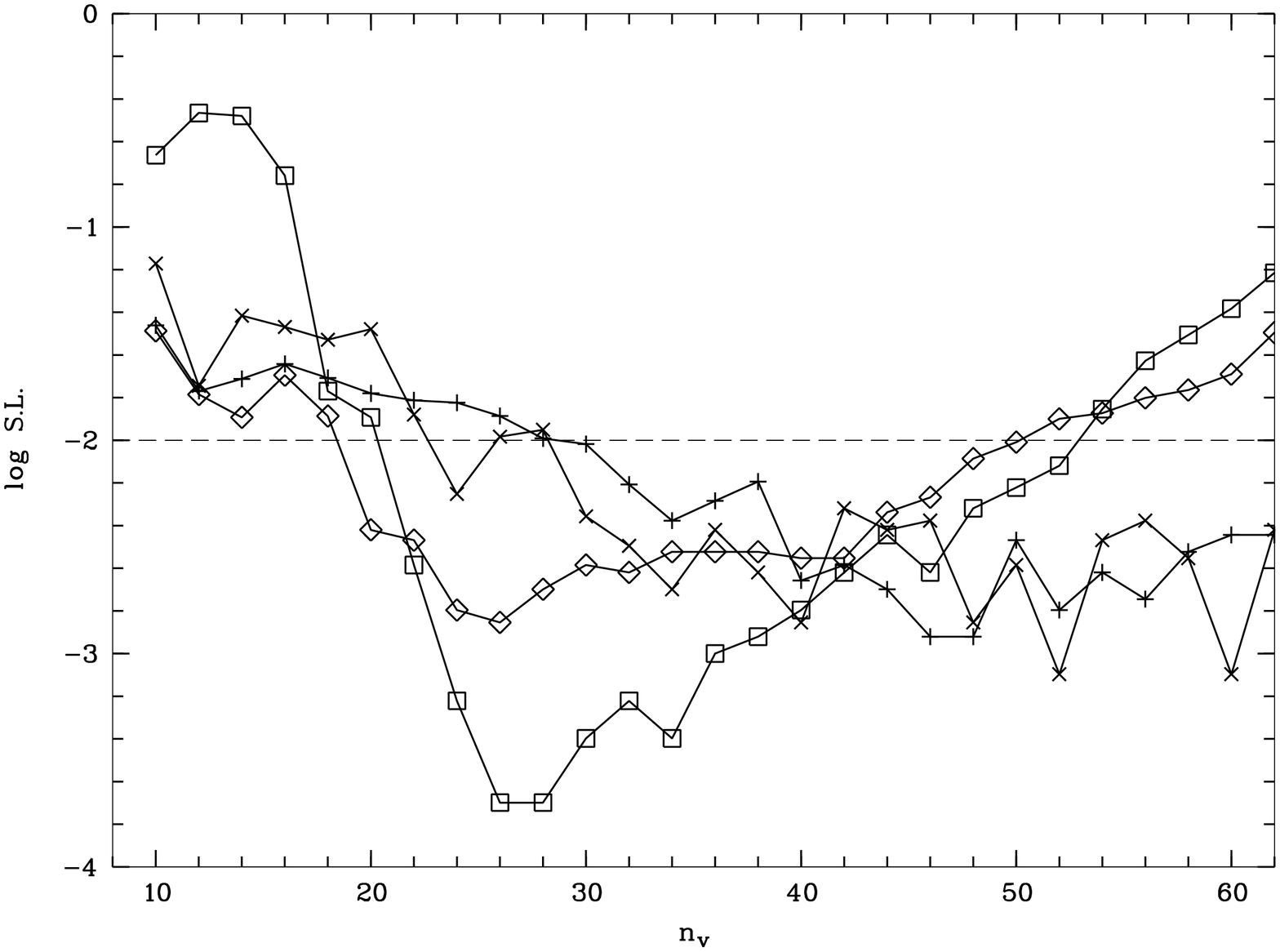}}
\caption[ ]{The logarithmic significance level (S.L.) of the four
statistical tests defined in Paper I, $S$ with $\Delta\theta_{c} =
60\degr$ $(\sq)$, $S_D (\diamondsuit)$, $Z_{c}^{m} (+)$, $Z_{c}
(\times)$, when applied to the new sample of 213 polarized quasars.
$n_v$ is the number of nearest neighbours around each quasar; it is
involved in the calculation of averaged quantities. {\bft The dashed
horizontal line indicates the 1\% significance level}}
\label{sl}
\end{figure}

Since in total 43 new polarized objects were found all over the sky,
it is also interesting to re-run the global statistical tests used in
Paper~I.  These tests are applied to the whole sample of 213
objects. The statistics basically measure the dispersion of
polarization position angles for groups of $n_v$ neighbours in the
3-dimensional space, the significance being evaluated through
Monte-Carlo simulations, shuffling angles over positions.  It is not
our purpose to repeat here what was done in Paper~I, but only to
illustrate the trend with the larger sample. {\bft The significance
levels of the statistical tests, i.e. the probabilities that the test
statistics would have been exceeded by chance only, are given in
Fig.~2 for the four tests considered in Paper I.}  Compared to Figs.~9
and 10 of Paper~I, all the statistical tests indicate a net decrease
of the significance level for the larger sample, strengthening the
view that polarization vectors are not randomly distributed over the
sky but are coherently oriented {\bft in groups of 20-30 objects}.  We
note a shift of the minimum significance level towards slightly higher
values of $n_v$, as expected from the increase of the number density
of the objects.

All these results confirm the existence of orientation effects in the 
distribution of quasar polarization vectors, and more particularly in 
the high-redshift region A1 where an independent test was performed.

\section{Towards an interpretation}

\subsection{Observational constraints}

First, it is important to note that the observational facts discussed
in detail in Paper~I and arguing against an instrumental bias and/or a
contamination by interstellar polarization in our Galaxy are still
valid. They are even strengthened since additional objects have been
observed, several of them by different authors with different
instrumentations. {\bft For the new measurements presented here, the
polarization of field objects has been measured simultaneously and
found to be very small (Lamy \& Hutsem\'ekers 2000).} The comparison
of the quasar polarization position angles with those of neighbouring
stars has also been re-investigated for the new sample. Using the most
recent compilation of Heiles (2000), we confirm the absence of
correlation between quasar and Galactic star polarization angles,
especially in region A1. And finally, the fact that the polarization
vectors of quasars on the same line of sight but at lower redshifts
are not accordingly aligned certainly remains one of the strongest
arguments against artifacts.

Let us now discuss some observational results providing us with
possible constraints on the phenomenon. {\bft This discussion is
mostly based on quasars in region A1.}
 
Looking at Fig.~1, we found --by chance-- that the plane of the Local
Supercluster (in the direction of its center) roughly passes through
the structure formed by the aligned objects. We have then transformed
the polarization position angles in the supergalactic coordinate
system (de Vaucouleurs et al.  1991) using Eq.~16 of Paper~I. A map is
illustrated in Fig.~3. It shows a rough alignment of quasar
polarization vectors with the supergalactic plane, the effect being
more prominent for those objects close to the supergalactic
equator. The most polarized objects (with $p \geq $ 5\%) do follow the
trend. This behavior is very reminiscent of the alignment of Galactic
star polarization vectors with the Galactic plane (Mathewson \& Ford
1970, Axon \& Ellis 1976).

\begin{figure}[t]
\resizebox{\hsize}{!}{\includegraphics*{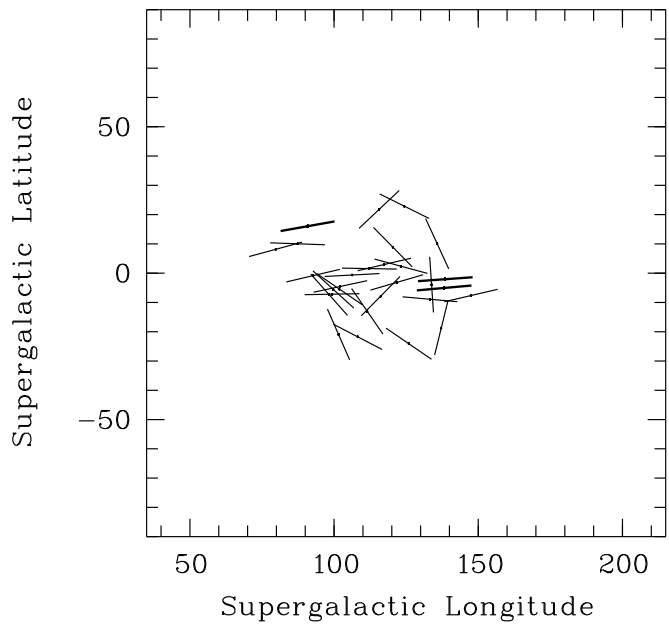}}
\caption[ ]{A map in the supergalactic coordinate system of the
polarization vectors of the 29 polarized quasars belonging to region
A1. The vector length is arbitrary. Thicker lines refer to objects
with $p \geq 5\%$}
\label{map3}
\end{figure}

Other constraints come from the relation between quasar intrinsic
properties and polarization. In Table~2, we give all quasars in region
A1 with good polarization measurements, either polarized ($p \geq 0.6
\%$ and $\sigma_{\theta} \leq 14\degr$ the latter constraint being
equivalent to $p / \sigma_{p} \simeq 2$), or unpolarized ($p < 0.6\%$
with $\sigma_{p} \leq 0.3\%$). The quasar type is also given: broad
absorption line (BAL), radio-loud non-BAL (RL), radio-quiet non-BAL
(RQ), or optically selected non-BAL (O). We may first notice that the
quasars with aligned polarization vectors do belong to all types, i.e.
radio-quiet / optically selected (3 objects), radio-loud (9), or BAL
(13). Note however that significantly polarized radio-quiet non-BAL
quasars are definitely less numerous, and that two of them,
B1115$+$080 and B1429$-$008, are possibly gravitationally
lensed. Furthermore, we can see that the known polarization difference
between BAL and non-BAL radio-quiet quasars also prevails in region A1
(Fig.~4). {\bft A Kolmogorov-Smirnov test gives a probability of only
0.6\% that these two samples of are drawn from the same parent
population}. The illustrated distributions are also in good agreement
with those reported by Hutsem\'ekers et al. (1998).  In addition, the
distribution of $p$ for non-BAL radio-quiet quasars is very similar to
that found by Berriman et al.  (1991) for the Palomar-Green quasar
sample. This clearly indicates that, also in region A1, quasar
polarization is related to the intrinsic properties of the objects.

\begin{table}
\caption[ ]{Polarized and unpolarized quasars in region A1}
\begin{tabular}{llrlrrrl}
\hline \\[-0.10in]%
Object  & $z$ & \  $p$ \ \ \ & \ $\sigma_{p}$ & $\theta$ \ & $\sigma_{\theta}$ & Ref & Type\\
(B1950) &     & \ \ \ (\%) \ &   (\%)         & ($\degr$)  & ($\degr$)         &     &     \\[0.05in] 
\hline \\[-0.10in]%
B1115$+$080 & 1.722 &   0.68 & 0.27 &   46 &  12 &  0 & RQ   \\
B1120$+$019 & 1.465 &   1.95 & 0.27 &    9 &   4 &  0 & BAL  \\
B1127$-$145 & 1.187 &   1.26 & 0.44 &   23 &  10 &  2 & RL   \\
B1138$-$014 & 1.270 &   0.38 & 0.23 &   53 &  21 &  9 & BAL  \\
B1138$+$040 & 1.876 &   0.10 & 0.24 &   36 &   - &  1 & RQ   \\
B1157$-$239 & 2.100 &   1.33 & 0.17 &   95 &   4 &  9 & BAL  \\
B1157$+$014 & 1.990 &   0.76 & 0.18 &   39 &   7 &  9 & RL   \\
B1158$+$007 & 1.380 &   0.44 & 0.20 &   74 &  14 &  9 & RL   \\
B1203$+$155 & 1.630 &   1.54 & 0.20 &   30 &   4 &  9 & BAL  \\
B1205$+$146 & 1.640 &   0.83 & 0.18 &  161 &   6 &  9 & BAL  \\
B1206$+$459 & 1.158 &   0.24 & 0.17 &  132 &  20 &  1 & RQ   \\
B1210$+$197 & 1.240 &   0.33 & 0.19 &   76 &  19 &  9 & RL   \\
B1212$+$147 & 1.621 &   1.45 & 0.30 &   24 &   6 &  0 & BAL  \\
B1215$+$127 & 2.080 &   0.62 & 0.24 &   17 &  12 &  9 & BAL  \\
B1216$+$110 & 1.620 &   0.58 & 0.19 &   63 &  10 &  9 & BAL  \\
B1219$+$127 & 1.310 &   0.68 & 0.20 &  151 &   9 &  9 & BAL  \\
B1222$-$016 & 2.040 &   0.80 & 0.22 &  119 &   8 &  9 & BAL  \\
B1222$+$146 & 1.550 &   0.23 & 0.18 &   65 &  30 &  9 & O    \\
B1222$+$228 & 2.046 &   0.84 & 0.24 &  150 &   8 &  2 & RL   \\
B1225$+$317 & 2.200 &   0.16 & 0.24 &  150 &   - &  2 & RL   \\
B1228$+$122 & 1.410 &   0.12 & 0.18 &  142 &   - &  9 & BAL  \\
B1230$-$237 & 1.840 &   0.05 & 0.19 &   72 &   - &  9 & O    \\
B1230$+$170 & 1.420 &   0.30 & 0.19 &  101 &  22 &  9 & BAL  \\
B1234$-$021 & 1.620 &   0.54 & 0.18 &   72 &  10 &  9 & O    \\
B1235$-$182 & 2.190 &   1.02 & 0.18 &  171 &   5 &  9 & RL   \\
B1238$-$097 & 2.090 &   0.18 & 0.18 &   80 &   - &  9 & O    \\
B1239$+$099 & 2.010 &   0.82 & 0.18 &  161 &   6 &  9 & BAL  \\
B1239$+$145 & 1.950 &   0.18 & 0.20 &   21 &   - &  9 & RQ   \\
B1241$+$176 & 1.273 &   0.12 & 0.19 &  120 &   - &  1 & RL   \\
B1242$+$001 & 2.080 &   0.22 & 0.19 &   56 &  39 &  9 & RQ   \\
B1246$-$057 & 2.236 &   1.96 & 0.18 &  149 &   3 &  8 & BAL  \\
B1246$+$377 & 1.241 &   1.71 & 0.58 &  152 &  10 &  2 & RL   \\
B1247$+$267 & 2.038 &   0.41 & 0.18 &   97 &  12 &  1 & RQ   \\
B1248$+$401 & 1.030 &   0.20 & 0.19 &    3 &   - &  1 & RQ   \\
B1250$+$012 & 1.690 &   0.21 & 0.18 &  132 &  37 &  9 & BAL  \\
B1254$+$047 & 1.024 &   1.22 & 0.15 &  165 &   3 &  1 & BAL  \\
B1255$-$316 & 1.924 &   2.20 & 1.00 &  153 &  12 &  4 & RL   \\
B1256$-$220 & 1.306 &   5.20 & 0.80 &  160 &   4 &  7 & RL   \\
B1256$-$175 & 2.060 &   0.91 & 0.19 &   71 &   6 &  9 & RL   \\
B1258$-$164 & 1.710 &   0.53 & 0.18 &  132 &  10 &  9 & O    \\
B1303$+$308 & 1.770 &   1.12 & 0.56 &  170 &  14 &  3 & BAL  \\
B1305$+$001 & 2.110 &   0.70 & 0.22 &  151 &   9 &  9 & O    \\
B1309$-$216 & 1.491 &  12.30 & 0.90 &  160 &   2 &  4 & RL   \\
B1309$-$056 & 2.212 &   0.78 & 0.28 &  179 &  11 &  0 & BAL  \\
B1317$+$277 & 1.022 &   0.15 & 0.20 &   94 &   - &  2 & RQ   \\
B1329$+$412 & 1.930 &   0.36 & 0.21 &   83 &  16 &  1 & RQ   \\
B1331$-$011 & 1.867 &   1.88 & 0.31 &   29 &   5 &  0 & BAL  \\
B1333$+$286 & 1.910 &   5.88 & 0.20 &  161 &   1 &  9 & BAL  \\
B1334$+$262 & 1.880 &   0.23 & 0.19 &  116 &  34 &  9 & BAL  \\
B1338$+$416 & 1.219 &   0.37 & 0.19 &   67 &  15 &  1 & RQ   \\
B1354$-$152 & 1.890 &   1.40 & 0.50 &   46 &  10 &  4 & RL   \\
B1416$+$067 & 1.439 &   0.77 & 0.39 &  123 &  14 &  2 & RL   \\
B1429$-$008 & 2.084 &   1.00 & 0.29 &    9 &   9 &  0 & RQ   \\
B1429$-$006 & 1.180 &   0.07 & 0.20 &  107 &   - &  9 & BAL  \\
\hline\end{tabular}
\footnotesize{References: 
(0) Hutsem\'ekers et al. 1998, (1) Berriman et al. 1990, 
(2) Stockman et al. 1984, (3) Moore \& Stockman 1984, 
(4) Impey \& Tapia 1990, (7) Visvanathan \& Wills 1998, 
(8) Schmidt \& Hines 1999, (9) Lamy \& Hutsem\'ekers 2000}
\end{table}

\subsection{Discussion}

\begin{figure}[t]
\resizebox{\hsize}{!}{\includegraphics*{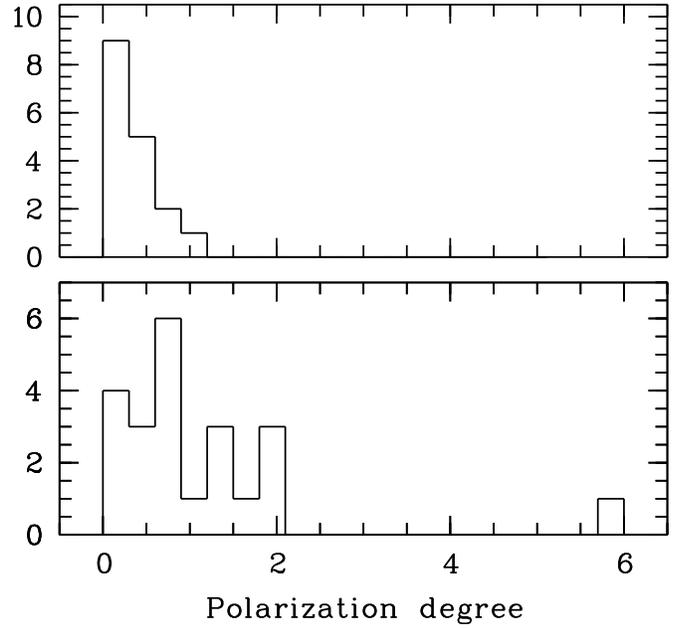}}
\caption[ ]{The distribution of the polarization degree $p$ (in \%)
for the quasars located in the region of alignment A1 (Table~2). 
Upper panel: radio-quiet (RQ+O) quasars. Lower panel: BAL quasars}
\label{hist1}
\end{figure}

The apparent alignment of quasar polarization vectors with the
supergalactic plane is very appealing as the starting point of an
explanation, namely since this could decrease by more than one order
of magnitude the scale at which a mechanism must act coherently. By
analogy with the alignment of stellar polarization vectors with the
plane of our Galaxy (Mathewson \& Ford 1970, Axon \& Ellis 1976), some
dichroism could be achieved due to extinction by dust grains aligned
in a magnetic field. Another possibility could be the conversion
of photons into pseudo-scalars also within a magnetic field (Harari \&
Sikivie 1992, Gnedin \& Krasnikov 1992, Gnedin 1994).  In both cases
the hypothetical magnetic field should be coherent on a $\sim$50 Mpc
scale, which is only slightly larger than the large-scale magnetic
field possibly detected by Vall\'ee (1990) in the direction of the
Virgo cluster. But this interpretation suffers some drawbacks: namely,
it cannot explain why the polarization vectors of quasars at lower
redshifts are not accordingly aligned (cf. Fig.~5 in Paper~I).

\begin{figure}[t]
\resizebox{\hsize}{!}{\includegraphics*{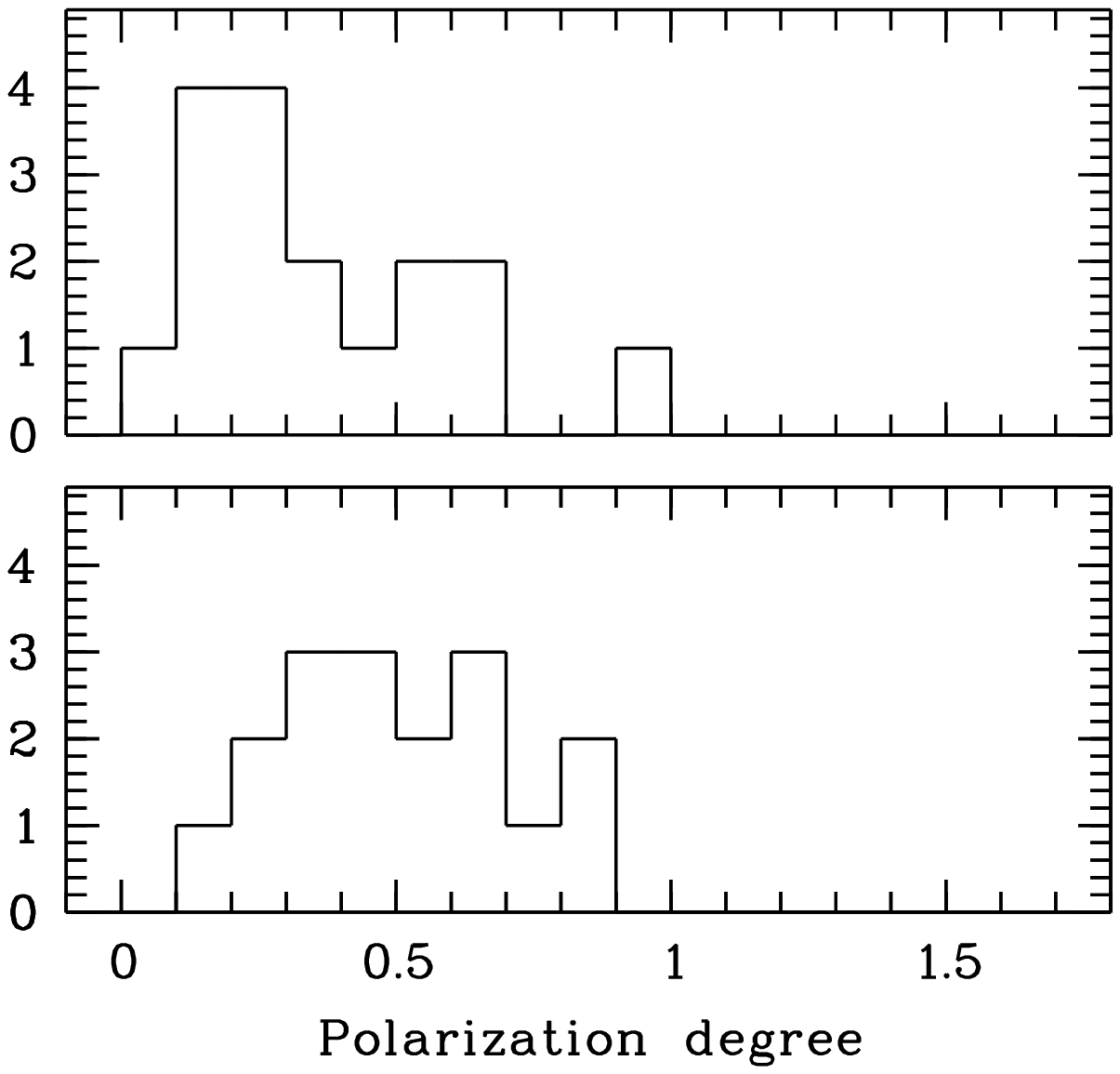}}
\caption[ ]{The distribution of the polarization degree for the
radio-quiet (RQ+O) quasars in region A1. The upper panel is an
enlargement of that of Fig.~4 and refers to the observed polarization
degree $p$ (in~\%).  The so-called polarization bias (due to the fact 
that $p$ is always a positive quantity) affects the first bin (see also
Berriman et al. 1991).  The lower panel shows the distribution of the
polarization degree after subtraction of a small but systematic
polarization ($p_s =0.25\%$, $\theta_s = 170\degr$) }
\label{hist2}
\end{figure}

It is therefore difficult to escape the conclusion that if a mechanism
is able to produce the alignment of polarization vectors of
high-redshift quasars by modifying the polarization state of photons
during their travel towards us, this must happen at redshifts $z \ga
1$ (for region A1), or all along the line of sight assuming a
cumulative or oscillatory effect (see also discussion in
Paper~I). Interestingly, an oscillation of quasar polarization with
cosmological distance has been predicted as the consequence of the
conversion of photons into pseudo-scalars within a large-scale
magnetic field permeating the intergalactic medium (Harari \& Sikivie
1992, Gnedin \& Krasnikov 1992, Gnedin 1994). However, this
interpretation, like other mechanisms which affect light as it
propagates towards us, cannot easily explain why correlations observed
between quasar polarization and intrinsic properties are not washed
away.

In this view we may ask ourselves how small could be the systematic
polarization which, added to randomly oriented polarization vectors,
can be at the origin of an orientation effect which involves nearly
all polarized quasars of region A1. To simulate this, we have
vectorially subtracted a systematic polarization $p_s$ oriented at
$\theta_s = 170\degr$ (the dominant direction, cf. Fig.~1) from all
polarized and unpolarized quasars of region A1 (Table~2). {\bft Then
we have re-selected a sample of significantly polarized quasars with
the conditions $p\geq 0.6\%$ and $\sigma_{\theta}\leq 14\degr$. For
$p_s = 0.25\%$, 29 quasars fulfil the conditions\footnote{\bft The fact
that this modified sample contains 29 objects exactly as the original
one is only chance. Several objects are indeed different} and only 15
objects out of this new sample have their polarization position angles
still in the range $146\degr - 46\degr$.} This small systematic
polarization seems therefore sufficient to produce the orientation
effect, and small enough to preserve the difference between BAL and
non-BAL quasars. However, as seen in Fig.~5, even such a small
systematic polarization significantly modifies the polarization
distribution of non-BAL radio-quiet quasars {\bft which appears
depleted at polarization degrees $p \la p_s$, and then} quite
different from the distribution found by Berriman et al. (1991) and
Hutsem\'ekers et al (1998). If we further note that values of $p_s$
higher than 0.25\% are obviously needed to explain the alignment of
the 16 objects with $p \geq 1.0\%$, we may conclude that it is quite
difficult to invoke a systematic additional polarization along the
line of sight without modifying the quasar intrinsic polarization
properties. This simple test also provides additional evidence that a
systematic instrumental effect is unlikely.

{\bft Although more subtle and speculative effects modifying the
polarization of light along the line of sight can probably be
imagined,} we may admit on the other hand that the quasars themselves
i.e. their structural axes are coherently oriented on Gpc scales. For
radio-loud quasars, it is well known that the optical polarization is
often parallel to the structural axis of the radio core (Rusk 1990,
Impey et al. 1991). Unfortunately, only one polarized quasar in region
A1 (B1127$-$145) is spatially resolved. It is however very interesting
to note that this object has a core structure parallel to its
polarization vector (Impey \& Tapia 1990). In this view it is worth
noting that possible coherence of morphological structures from the
central engines of active galactic nuclei to superclusters has been
suggested by West (1991, 1994), although this observation is
apparently not confirmed at the supercluster scale (Jaaniste et
al. 1998). From the theoretical point of view, some studies (Reinhardt
1971, Wasserman 1978) have pointed out the possible effects of
magnetic fields on galaxy formation and orientation. Extrapolating, a
correlation between quasar structural axes could be settled at the
epoch of formation and related to very large-scale primordial magnetic
fields possibly formed during inflation (Battaner \& Florido
2000). Large-scale vorticity is another possibility, at least
qualitatively. In this view, the apparent alignment found with the
supergalactic plane is puzzling. However, coincidence cannot be ruled
out, especially if we note that the polarization vectors of quasars
belonging to the other regions of alignments (regions A2 and A3 in
Paper~I) do not show the same alignment with the plane of the Local
Supercluster.

\section{Conclusions}

In order to verify the existence of very large-scale coherent
orientations of quasars polarization vectors, we have obtained new
polarization measurements for quasars located in a given region of the
sky where the range of polarization position angles was predicted in
advance. {\bft The statistical analysis of this new sample provides an
independent confirmation of the existence of alignments of quasar
polarization vectors at high redshifts.}  In total, out of 29
polarized quasars located in this region, 25 have their polarization
vectors coherently oriented.  {\bft Moreover, global statistical tests
applied to the whole sample indicate that, with the increased size of
the sample, the detection of the effect is stable and even more
significant.}

The increased size of the sample also allowed us to put some
constraints on the phenomenon. We namely found that the polarization
vectors of quasars in region A1 are apparently parallel to the plane
of the Local Supercluster, while those of quasars at lower redshifts
are not accordingly aligned. Furthermore, we found that the well-known
correlations between quasar intrinsic properties and polarization also
prevail for those quasars located in that region of alignment.

Mechanisms modifying the polarization somewhere along the line of
sight could explain some results, but cannot easily fit all
constraints, namely the fact that correlations between quasar
intrinsic properties and polarization are not washed away. Another
possibility is that the quasar structural axes themselves are
coherently oriented. However, given the scales involved, a reasonable
physical mechanism is far from obvious.  Interestingly, in region A1,
nine quasars with aligned polarization vectors are radio-loud, such
that this hypothesis could be tested by simply mapping their radio
core.

The interpretation of this orientation effect therefore remains
puzzling. Nevertheless, the presence of coherent orientations at such
large scales seems to indicate the existence of a new interesting
effect of cosmological importance.

\begin{acknowledgements}
We are grateful to the referee for useful comments.  The SIMBAD and
NED databases have been consulted namely to classify some quasars in
Table~2.
\end{acknowledgements}

\end{document}